\documentclass[aps,preprint,epsfig,rotate]{revtex4}
\begin{document}

\title{Properties of negatively charged lithium ions and evaluation of the half-life of ${}^{7}$Be atom(s)} 

 \author{Alexei M. Frolov}
 \email[E--mail address: ]{afrolov@uwo.ca}

 \affiliation{Department of Applied Mathematics, University of Western Ontario,
              London, Ontario, N6H 5E5, CANADA}

\author{David M. Wardlaw}
 \email[E--mail address: ]{dwardlaw@mun.ca}

\affiliation{Department of Chemistry, Memorial University of Newfoundland, St.John's, 
             Newfoundland and Labrador, A1C 5S7, Canada}

\date{\today}

\begin{abstract}

Bound state properties of the four-electron lithium ion Li$^{-}$ in its ground $2^{2}S-$state and isotope-subsistuted ${}^{6}$Li$^{-}$ and ${}^{7}$Li$^{-}$ ions in 
their ground $2^1S-$state(s) are determined from the results of accurate, variational computations. Another closely related problem discussed in this study is 
accurate numerical evaluation of the half-life of the beryllium-7 isotope. This problem is of paramount importance for modern radiochemistry.


\end{abstract}

\maketitle

\section{Introduction}

In this communication we consider the bound states properties of the negatively charged Li$^{-}$ ion in its ground $2^1S(L = 0)-$state, or $2^1S-$state, for short. It is well 
known that such an ion has only one bound state which is the ground $2^{1}S-$state. The electronic structure of this state in the Li$^{-}$ ion corresponds to the $1s^2 2s^2$ 
electron configuration. Recently, the negatively charged lithium ion Li$^{-}$ has become of interest in numerous applications, since formation of these ions is an important step 
for workability of the compact lithium and/or lithium-ion electric batteries. Both lithium and lithium-ion batteries are very compact, relatively cheap and reliable sources of 
constant electric current which are widely used in our everyday life and in many branches of modern industry. However, it appears that the Li$^{-}$ ion is not a well studied 
atomic system. Indeed, many bound state properties of this ion have not been evaluated at all, or were known only approximately. Moreover, those properties that have been evaluated 
are for an isolated Li$^{-}$ ion in a vacuum. In reality, it is crucial to evaluate the bound state properties in the presence of different organic acids which are extensively used 
in lithium-ion batteries. 

Our goal in this study is to determine various bound state properties of the four-electron (or five-body) Li$^{-}$ ion. It should be mentioned that most of these properties have 
not been evaluated in earlier studies. To lowest-order approximation in the fine structure constant $\alpha (= \frac{e^2}{\hbar c}$), the negatively charged Li$^{-}$ ion 
is described by the non-relativistic Schr\"{o}dinger equation $H \Psi = E \Psi$, where $H$ is the non-relativistic Hamiltonian and $E(< 0)$ is the eigenvalue. Without loss of 
generality we shall assume that the bound state wave function $\Psi$ has the unit norm. The non-relativistic Hamiltonian $H$ of an arbitrary four-electron atom/ion takes the form 
(see, e.g., \cite{LLQ})
\begin{eqnarray}
 H = -\frac{\hbar^2}{2 m_e} \Bigl[\nabla^2_1 + \nabla^2_2 + \nabla^2_3 + \nabla^2_4 + \frac{m_e}{M} \nabla^2_5 \Bigr] - Q e^2 \sum^{4}_{i=1} \frac{1}{r_{15}} 
 + e^{2} \sum^{3}_{i=1} \sum^{4}_{j=2 (j>i)} \frac{1}{r_{ij}} \; \; \; , \; \; \label{Hamil}
\end{eqnarray}
where $\hbar = \frac{h}{2 \pi}$ is the reduced Planck constant, $m_e$ is the electron mass and $e$ is the electric charge of an electron. In this equation and everywhere below in 
this study the subscripts 1, 2, 3, 4 designate the four atomic electrons $e^-$, while the subscript 5 (= $N$) denotes the heavy atomic nucleus with the mass $M$ ($M \gg m_e$), and 
the positive electric (nuclear) charge is $Q e$. The notation $r_{ij} = \mid {\bf r}_i - {\bf r}_j \mid = r_{ji}$ in Eq.(\ref{Hamil}) and everywhere below stands for the interparticle 
distances between particles $i$ and $j$. These distances are also called the relative coordinates to emphasize their differences with the three-dimensional coordinates ${\bf r}_i$, 
which are the Cartesian coordinates of the particle $i$. In Eq.(\ref{Hamil}) and everywhere below in this work we shall assume that $(ij)$ = $(ji)$ = (12), (13), (14), (15), (23), 
(24), (25), (34), (35) and (45), for four-electron atomic systems and particle 5 means the atomic nucleus. Analogously, for three-electron atomic systems we have $(ij)$ = $(ji)$ = 
(12), (13), (14), (23), (24) and (34), where particle 4 is the atomic nucleus. Below only atomic units $\hbar = 1, \mid e \mid = 1, m_e = 1$ are employed. In these units the 
explicit form of the Hamiltonian $H$, Eq.(\ref{Hamil}), is simplified and takes the form
\begin{eqnarray}
 H = -\frac12 \Bigl[\nabla^2_{1} + \nabla^2_{2} + \nabla^2_{3} + \nabla^2_{4} + \frac{m_e}{M} \nabla^2_{5} \Bigr] - Q \sum^{4}_{i=1} \frac{1}{r_{i5}} + \sum^{3}_{i=1} 
 \sum^{4}_{j=2 (j>i)} \frac{1}{r_{ij}} \; \; \; , \; \; \label{Hamil1}
\end{eqnarray}
where $Q$ is the nuclear charge of the central positively charged nucleus. For the neutral Li-atom and negatively charged Li$^{-}$ ion we have $Q = 3$. Note also that in this study the 
notations Li and Li$^{-}$ stand for the lithium atom and ion with the infinitely heavy nucleus, while analogous notations with the superscript 6 (and/or 7) mean the ${}^{6}$Li 
(${}^{7}$Li) atom and ${}^{6}$Li$^{-}$ (${}^{7}$Li$^{-}$) ion, respectively. The nuclear mass of the lithium-6 nucleus used in this study is $M({}^{6}$Li) = 10961.8968 $m_e$, while the 
nuclear mass of the lithium-7 nucleus is $M({}^{7}$Li) = 12786.3927 $m_e$.

As mentioned above the bound state spectrum of the negatively charged Li$^{-}$ ion includes only one bound $2^1S-$state, which is the ground state. The stability of this state means 
stability against its dissociation (or ionization) Li$^{-}$ = Li$(2^2S)$ + $e^{-}$, where the notation Li$(2^2S)$ means the lithium atom in its ground (doublet) $2^2S-$state. 
Stability of this state in the Li$^{-}$ ion has been known since the first accurate calculations performed the middle of 1970's. Note that all methods based on the Hartree-Fock 
approximation cannot produce any bound state in the Li$^{-}$ ion. Furthermore, it is clear that to determine the ground (bound) $2^1S-$state in the Li$^{-}$ ion such an variational 
expansion must be truly correlated and accurate. Our method discussed in the next Section allows one to construct accurate and highly accurate wave functions for arbitrary few-body 
systems. This includes few-electron atomic systems, e.g., the Li$^{-}$ ion, Be atom and Li atom. By using this method we determine a large number of bound state properties of the 
four-electron Li$^{-}$ ion(s) and Be atom(s). The same method is applied to obtain highly accurate wave function(s) of the ground $2^2S-$state in the three-electron Li atom(s). It is very 
interesting to compare the bound state properties of the ground $2^1S-$state in the Li$^{-}$ ion and the ground $2^2S-$state in the neutral Li atom. Another unsolved problem known for 
four-electron atomic systems is to explain experimental variations of the half-life of the ${}^{7}$Be isotope in different chemical compounds, or molecules. This problem can be solved by 
applying the same computational methods which we used for the Li$^{-}$ ion and is considered in the third Section. Concluding remarks can be found in the last Section.

\section{Variational wave functions}

To determine accurate solutions of the non-relativistic Schr\"{o}dinger equation $H \Psi = E \Psi$ in this study we approximate the unknown wave function(s) by a variational expansion in 
multi-dimensional gaussoids. Each of these gaussoids explicitly depends upon relative coordinates $r_{ij}$. There are ten such coordinates for the five-body (or four-electron) atomic 
problem and six coordinates for the four-body (or three-electron) atomic problem. For the singlet ${}^{1}S(L = 0)-$states in four-electron atomic systems this expansion takes the form 
\cite{KT}, \cite{Fro1}:
\begin{eqnarray}
  \psi(L = 0; S = 0) = \sum^{N_A}_{i=1} C_i \exp(-\alpha_{ij} r^2_{ij}) \chi^{(1)}_{S=0} + \sum^{N_B}_{i=1} G_i \exp(-\beta_{ij} r^2_{ij}) \chi^{(2)}_{S=0} \; \; \; \label{equa1}
\end{eqnarray}
where $C_i$ (and $G_i$) are the linear variational coefficients of the trial wave function, while $\alpha_{ij}$ are the ten non-linear parameters and (ij) = (12), (13), 
$\ldots$, (45). Analogously, the notation $\beta_{ij}$ stands for other ten non-linear parameters which must also be varied (independently of $\alpha_{ij}$) in calculations. 
Notations $\chi^{(1)}_{S=0}$ and $\chi^{(2)}_{S=0}$ in Eq.(\ref{equa1}) designate the two independent spin functions which can be considered for the singlet $2^1S-$state, or 
$(2^{1}S \mid 1s^2 2s^2)-$electron configuration. The explicit forms of these two spin functions are:
\begin{eqnarray}
  \chi^{(1)}_{S=0} &=& \alpha \beta \alpha \beta + \beta \alpha \beta \alpha - \beta \alpha \alpha \beta - \alpha \beta \beta \alpha \;\;\; \label{spin1} \\
 \chi^{(2)}_{S=0} &=& 2 \alpha \alpha \beta \beta + 2 \beta \beta \alpha \alpha - \beta \alpha \alpha \beta - \alpha \beta \beta \alpha - \beta \alpha \beta \alpha 
  - \alpha \beta \alpha \beta \;\;\; \label{spin2} 
\end{eqnarray}
In numerical calculations of the total energies and other spin-independent properties (i.e. expectation values) one can always use just one spin function, e.g., $\chi^{(1)}_{S=0}$ 
from Eq.(\ref{spin1}). It follows from the fact that the Hamiltonian Eq.(\ref{Hamil1}) does not depend explicitly upon the electron spin and/or any of its components. 

The radial basis functions in Eq.(\ref{equa1}) are the multi-dimensional gaussoids, or five-dimensional gaussoids (see, Eq.(\ref{equa1}) above). These basis functions are not orthogonal 
to each other. Therefore, the original Schr\"{o}dinger equation $H \Psi = E \Psi$ is reduced to the solution of the following eigenvalue problem $\hat{H} {\bf C} = E \hat{S} {\bf C}$, 
where $\hat{H}$ and $\hat{S}$ are the matrixes of the Hamiltonian and overlap, respectively, while $E$ is the unknown eigenvalue and ${\bf C}$ is the vector formed from the linear 
variational coefficients $C_i$ mentioned in Eq.(\ref{equa1}). It is straightforward to derive analogous equations in those cases where the corresponding vector also includes non-zero 
${\bf G}$-components (see Eq.(\ref{equa1})). In fact, the linear coefficients $C_i$ and $G_i$ from Eq.(\ref{equa1}) can be considered as components of the $N_A-$ and $N_B-$dimensional 
vectors ${\bf C}$ and ${\bf G}$, respectively. Now, it is clear that we need to obtain the explicit formulas for all matrix elements of the Hamiltonian and overlap matrix. Note that this
part of the problem was solved long ago in \cite{KT}. Formulas which include additional complications related with the spin parts of the wave functions for four- and three-electron atomic 
systems have been derived and presented in a number of our papers (see, e.g., \cite{Fro2}). Therefore, below we restrict ourselves to a brief description of this method.

First, we construct the wave functions of the correct permutation symmetry by applying the complete anti-symmetrizer ${\cal A}_{1234}$ to the trial wave function represented in the form 
of Eq.(\ref{equa1}). As analytical expression for ${\cal A}_{1234} \psi(L = 0; S)$ is represented as a finite sum of different spatial and spin terms. In each term arguments in the spin 
and spatial functions are interchanged. At the second step of the procedure we need to calculate (analytically) the overlap integrals between all components of the spin functions with 
the original and interchanged arguments, after which we obtain the final expression for the matrix elements of the Hamiltonian $\hat{H}$ and overlap $\hat{S}$ matrixes. In reality, all 
matrix elements of $\hat{H}$ and $\hat{S}$ contain finite sums of the radial matrix elements with some numerical (integer) coefficients which are determined by the overlap integrals of 
the spin functions (for more details, see \cite{Fro2}). The procedure used to solve three-electron atomic problems (bound states) is completely analogous and here we do not want to repeat 
its description (see, e.g., \cite{Fro2}). Note also that below we apply three-electron version of our procedure to determine the bound state properties of the ground $2^2S-$state in the 
neutral Li-atom.

As mentioned above in this paper we consider the ground $2^1S-$state of the Li$^{-}$ ion (or ${}^{\infty}$Li$^{-}$ ion). Our goal is to determine the total energy of this state and 
expectation values of some of its properties. Such properties include a few powers of interparticle distances $\langle r^{n}_{ij} \rangle$, where $n = -2, -1, 1, 2, 3, 4$ (for $n = 0$ each 
of these expectation values equals unity), electron-nucleus and electron-electron delta-functions, single electron kinetic energy $\langle \frac12 {\bf p}^2_e \rangle$, electron-nucleus and 
electron-electron kinetic correlations $\langle {\bf p}_e \cdot {\bf p}_N \rangle$ and $\langle {\bf p}_e \cdot {\bf p}_e \rangle$ and a few others (see also discussion in the Appendix). 
Numerical values of these expectation values can be found in Table I. Table I contains analogous properties of the ${}^{6}$Li$^{-}$ and ${}^{7}$Li$^{-}$ ions which also have only one stable 
$2^{1}S-$state. In addition to these ions, Table I includes the bound state properties of the ground $2^{2}S-$state in the neutral Li atom, which is a three-electron atomic system. All these 
properties are expressed in atomic units. The expectation values of the different bound state properties computed for the ground $2^2S-$state of the three-electron Li atom (or ${}^{\infty}$Li 
atom) are of interest to make a direct comparison with the analogous properties of the four-electron Li$^{-}$ ion (or ${}^{\infty}$Li$^{-}$ ion). It should be mentioned that our method allows 
one to determine the total energies of the bound states in the four- and three-electron atomic systems to relatively high accuracy. For the ${}^{\infty}$Li$^{-}$ ion our energy ($E$ = 
-7.50076754 $a.U.$) is one of the best total energies ever computed for this ion. In fact, the total energy of the ground state in the Li atom (or ${}^{\infty}$Li atom) is  comparable to the 
accuracy known for this atom from Hy-CI calculations with 1000 - 2000 terms in the wave function. The total energies of the $2^2S-$state in the Li atom determined with $N = N_A$ = 700 and $N 
= N_A$ = 1000 in Eq.(\ref{equa1}) are $E$ = -7.478060003 $a.u.$ and $E$ = -7.478060107 $a.u.$, respectively. These values are close to the current `exact' total energy -7.4780603236503 $a.u.$, 
obtained with the use of many thousands of Hylleraas (Hy) and Hylleraas-CI basis functions. Note that our energies for the ground states in both Li$^{-}$ ion and Li atom are still converging 
and we hope soon to report the updated value of the total energy which has significantly better overall accuracy.  
  
Let us compare the bound state properties of the Li$^{-}$ ion in the ground $2^{1}S-$state with the analogous properties of the neutral Li atom in its ground $2^2S-$state. First, we note some 
substantial differences in the electron-nucleus and electron-electron distances $\langle r_{eN} \rangle$ and $\langle r_{ee} \rangle$. For the Li$^{-}$ ion these distances are larger 
(significantly larger) than for the neutral Li atom. The same statement is correct for all positive powers of these inter-particle distances, i.e. for the $\langle r^{k}_{eN} \rangle$ and 
$\langle r^{k}_{ee} \rangle$ (where $k$ is integer and $k \ge 2$) expectation values shown in Table I. This is an indication of the known fact that the Li$^{-}$ ion is a weakly-bound, 
four-electron system atomic system. This fact can be confirmed by calculation of the following dimensionless ratio 
\begin{equation}
   \epsilon = \frac{E({\rm Li}^{-}) - E({\rm Li})}{E({\rm Li}^{-})} \approx 0.00301
\end{equation}
where $E($Li$^{-})$ and $E($Li) are the total energies of the negatively charged Li$^{-}$ ion in the ground $2^{1}S-$state and Li atom in the ground $2^2S-$state. A very small value of this 
parameter $\epsilon$, which here is significantly less that 0.01 (or 1 \%), is a strong indication that the Li$^{-}$ ion is an extremely weakly-bound atomic system. This allows 
one to represent the internal structure of the bound state in the Li$^{-}$ ion as a motion of one electron in the `central' field created by the infinitely heavy Li atom. In other words, 
the electronic structure of this ion is $1s^2 2s^2$ and one of the two outer-most electrons moves at very large distances from the central nucleus. In reality, this representation 
is only approximate, since, e.g., there is an exchange symmetry between two electrons form the $2s^2$ shell. Nevertheless, such a `cluster' structure can be useful to predict and 
explain a large number of bound state properties of the Li$^{-}$ ion. For instance, consider the expectation value of the inverse electron-nucleus distance, i.e. $\langle r^{-1}_{eN}
\rangle$. From the definition of this expectation value we write the following expression
\begin{equation}
  \langle r^{-1}_{eN} \rangle = \frac{1}{4} \Bigl( \langle r^{-1}_{1N} \rangle + \langle r^{-1}_{2N} \rangle + \langle r^{-1}_{3N} \rangle + \langle r^{-1}_{4N} \rangle \Bigr)
  \label{eqsym1}
\end{equation}
where all expectation values in the right-hand side are determined without any additional symmetrization between four electrons. As mentioned above the Li$^{-}$ ion has a sharp cluster 
structure and its fourth electron is located on avarage far away from the central nucleus. This means that $\langle r^{-1}_{4N} \rangle \approx 0$. In this case it follows from Eq.(\ref{eqsym1}) 
that 
\begin{equation}
  \langle r^{-1}_{eN} \rangle = \frac{3}{4} \langle r^{-1}_{1N} \rangle = \frac{3}{4} \langle r^{-1}_{eN} \rangle \approx \frac{3}{4} \langle r^{-1}_{eN} \rangle_{{\rm Li}} \label{eqsym2}
\end{equation}
where $\langle r^{-1}_{eN} \rangle_{{\rm Li}}$ is the corresponding expectation value for the neutral Li-atom. It is clear that this equality is only approximate. Analogous approximate 
evaluations can be obtained for some other properties, e.g., for the expectation values of all delta-functions and inverse powers of electron-nucleus and electron-electron distances. 

Table I contains a large number of bound state properties of the negatively charged Li$^{-}$ ion. Numerical values of these properties are of interest in various scientific and technical 
applications, including quite a few applications to electro-chemistry of the lithium and lithium-ion batteries. Our expectation values form a complete basis set of numerical values which 
can be useful in analysis of different macroscopic systems containing lithium atoms and negatively charged ions.     

\section{On the half-life of the beryllium-7 isotope}

Results of our accurate computations of the ground $2^1S-$state in the weakly-bound Li$^{-}$¨ion indicate clearly that our variational expansion Eq.(\ref{equa1}) is very effective in 
applications to four-electron atomic systems. In this Section we apply the same variational expansion, Eq.(\ref{equa1}), to investigate another long-standing problem known in the atomic physics 
of four-electron atomic systems. Briefly, our goal is to explain variations of the half-life of the beryllium-7 isotope in different chemical enviroments. As follows from the results of 
numerous experiments the half-life of the ${}^{7}$Be isotope is `chemically dependent', i.e. it varies by $\approx$ 0.5 \% - 5 \% for different chemical compounds. This fact contradicts an old 
fundamental statement (see, e.g., \cite{Remi}) that actual decay rates of chemical isotopes cannot depend upon their chemical enviroments. This explains a substantial interest in chemical 
compounds which contain atoms of the beryllium-7 (or ${}^{7}$Be) isotope. It should be mentioned that in modern laboratories different chemical compounds containing ${}^{7}$Be atoms 
are not `exotic' substances, since the nuclei of ${}^{7}$Be are formed in the $(p;n)-$ and $(p;\alpha)-$reactions of the ${}^{7}$Li and ${}^{10}$B nuclei with the accelerated protons. A few other 
nuclear reactions involving nuclei of some light and intermediate elements, e.g., C, Al, Cu, Au, etc, also lead to the formation of the ${}^{7}$Be nuclei. In general, an 
isolated ${}^{7}$Be nucleus decays by using a few different channels, the most important of which is the electron capture (or $e^{-}-$capture) of one atomic electron from the 
internal $1s^2-$shell. The process is described by a simple atomic-nuclear equation ${}^{7}$Be $\rightarrow$ ${}^{7}$Li, where there is no free electron emitted after the process. During this 
process the maternal ${}^{7}$Be nucleus is transformed into the ${}^{7}$Li nucleus which can be found either in the ground, or in the first excited state. The following transition of the 
excited ${}^{7}$Li$^{*}$ nucleus into its ground state ${}^{7}$Li proceeds with the emission of a $\gamma-$quantum which has the energy $E_{\gamma} \approx 0.477$ $MeV$. Such $\gamma-$quanta 
can easily be registered in modern experiments and this explains numerous applications of chemical compounds of ${}^{7}$Be in radio-chemistry.      

Let us discuss the process of the electron capture in the ${}^{7}$Be-atom in detail. Assume for a moment that all ${}^{7}$Be atoms decay by the electron capture from the ground (atomic) 
$2^{1}S-$state. In this case, by using the expectation value of the electron-nucleus delta-function $\langle \delta({\bf r}_{eN}) \rangle$ computed for the ground $2^1{}S-$state of an 
isolated Be-atom we can write the following expression for the half-life $\tau$ of the ${}^{7}$Be atom/isotope
\begin{equation}
   \tau = \frac{1}{\Gamma} = \frac{1}{A \langle \delta({\bf r}_{eN}) \rangle} \; \; \; \label{eq1}
\end{equation}
where $\Gamma$ is the corresponding width and $A$ is an additional factor for a given compound of beryllium. The half-life $\tau$ determines the moment when 50 \% of the incident ${}^{7}$Be will 
decay by the electron capture. An analytical formula for $\tau$, Eq.(\ref{eq1}), follows from the fact that the corresponding width $\Gamma = \tau^{-1}$ must be proportional to the product of the
expectation value of the electron-nucleus delta-function and an additional factor $A$. The expectation value of the electron-nucleus delta-function computed with the non-relativistic 
wavefunction determines the electron density at the surface of a sphere with the spatial radius $R \approx \Lambda_e = \frac{\hbar}{m_e c} a_0 = \alpha a_0$, where $a_0$ is the Bohr radius $a_0 
\approx \frac{\hbar^2}{m_e e^2} (\approx 5.292 \cdot 10^{-9}$ $cm$), $c$ is the speed of light and $\Lambda_e$ is the Compton wave length. The `constant' $A$ in Eq.(\ref{eq1}) represents an 
`additional' probability for an electron (point particle) to penetrate from the distances $R \approx \Lambda_e = \alpha a_0$ to the surface of the nucleus $R_N \approx 1 \cdot 10^{-13}$ $cm$. 

Numerical value of $A$ can be evaluated by assuming that the mean half-life of the ${}^{7}$Be-atom in its ground $2^{1}S-$state equals 53.60 days and by using our best expectation value obtained 
for the expectation value of the electron-nucleus delta-function $\langle \delta({\bf r}_{eN}) \rangle \approx$ 8.82515 $a.u.$, one finds that $\Gamma \approx 2.1593422 \cdot 10^{-7}$ sec$^{-1}$. 
From here we find that the factor $A$ in Eq.(\ref{eq1}) equals
\begin{equation}
  A \approx \frac{2.1593422 \cdot 10^{-7}}{\langle \delta({\bf r}_{eN}) \rangle} \approx 2.439521 \cdot 10^{-8} \; \; \;  \label{eq2}
\end{equation}
where the expectation value $\langle \delta({\bf r}_{eN}) \rangle$ must be taken in atomic units. To move further we have to assume that the additional factor $A$ in Eq.(\ref{eq2}) does not depend 
either upon the conserved quantum numbers of the Be-atom, nor upon the actual chemical background of this atom. This is an approximation, but in actual applications the accuracy of this 
approximation is relatively high. In this case we can write the following formula for the ratio of half-life of the two different molecules X(Be) and Y(Be) which contain ${}^{7}$Be atoms
\begin{equation}
   \frac{\tau({\rm X(Be)})}{\tau({\rm Y(Be)})} = \frac{\langle \delta({\bf r}_{eN}); {\rm Y(Be)} \rangle}{\langle \delta({\bf r}_{eN}); {\rm X(Be)} \rangle} 
   \; \; \;  \label{eq3}
\end{equation}
Let us apply this formula to the case when one of the ${}^7$Be-atoms is in the ground $2^1S-$state, while another such an atom is in the triplet $2^{3}S-$state. The expectation value of the 
$\delta_{eN}$-function for the ground state in the Be-atom is given above, while for the triplet state we have $\langle \delta({\bf r}_{eN}) \rangle \approx$ 8.739558 $a.u.$ Both these expectation 
values were determined in our highly accurate computations of the ground $2^{1}S-$ and $2^{3}S-$state in the four-electron Be atom. With these numerical values one finds from Eq.(\ref{eq3}) that 
the half-life of the ${}^{7}$Be atom in its triplet $2^3S-$state is 1.009794 times (or by $\approx$ 1 \%) longer than the corresponding half-life 0f the ${}^{7}$Be atom in its ground singlet 
$2^1S$-state. This simple example includes two different bound states in an isolated ${}^{7}$Be-atom. In general, by using the formula Eq.(\ref{eq3}) we can approximately evaluate the half-life of 
the ${}^{7}$Be atoms in different molecules and compounds. The formula Eq.(\ref{eq3}) can be applied, e.g., to BeO, BeC$_2$, BeH$_2$ and many other beryllium compounds, including beryllium-hydrogen 
polymers, e.g., Be$_n$H$_{2n}$ for $n \approx 100 - 1000$ (see, e.g., \cite{Ref4} - \cite{Ref7} and references therein).

As is well known from atomic physics, the electronic structure of the excited bound states of thefour-electron  Be-atom(s) is $1s^2 2s n\ell$ (or $1s^2 2s^1 n\ell^1$), where $\ell \ge 0$ and $n \ge 3$. 
In general, such an excited state arises after excitation of a single electron from the $1s^2 2s^2$ electron configuration, which correspond to the ground state, or `core', for short. It is clear that 
the final $1s^2 2s^1 n\ell^1$ configuration is the result of a single electron excitation $2s \rightarrow n \ell$. All other states with excitation(s) of two and more electrons from the core are unbound. 
In general, a very substantial contribution ($\ge$ 95 \%) to the expectation value of the electron-nucleus delta-function comes from the two internal electrons (or $1s^2-$electrons) of the Be-atom.  
Briefly this means that the expectation value of the electron-nucleus delta-function is almost the same for all molecules which contain the bound Be-atom. Variations in 3 \% - 6 \% are possible and they
are related with the contribution of the  two outer-most electrons in the expectation value of the electron-nucleus delta-function $\langle \delta({\bf r}_{eN}) \rangle$. As follows from computational 
results the overall contribution from two outer-most electrons is only 3 \% - 6 \% of the total numerical value. This means that variations in the chemical enviroment of one ${}^{7}$Be atom can change the 
half-life of this atom by a factor of 1.03 to 1.06 (maximum). In reality, such changes are significantly smaller, but they can be noticed in modern experiments. 

It is interesting to note that analogous result (3 \% - 6 \% differences as maximum) can be predicted for other nuclear processes which are influenced by variations in the chemical enviroment, e.g., for 
the excitation of the ${}^{235}$U nucleus which also depends upon chemical  enviroment \cite{XX} - \cite{Fro2005}. It is well known (see, e.g., \cite{ZZ}) that the ${}^{235}$U nucleus 
has an excited state with the energy $\approx$ 75 - 77 $eV$. There is no such level in the ${}^{234}$U, ${}^{236}$U and ${}^{238}$U nuclei. Nuclear properties of the ground and first excited states in 
the ${}^{235}$U nucleus differ substantially. Moreover, by changing the actual chemical enviroment of the ${}^{235}$U atom we can change the probabilities of excitation of the central nucleus, e.g., by 
using different alloys of uranium, in order to change and even control nuclear properties. For instance, this approach can be used to achieve and even exceed critical conditions with respect to neutron 
fission. Theoretical evaluations and preliminary experiments show that possible changes in nuclear properties of different compounds of uranium-235 do not exceed 3 - 6 \%. It is very likely that 3 - 6 \% 
is the upper limit of influence of atomic (and molecular) properties on the nuclear properties of different isotopes. On the other hand, possible changes in atomic and molecular properties produced by 
processes, reactions and decays in atomic nuclei are always significant.    

Thus, if we know the expectation value of the electron-nucleus delta-function for the beryllium-7 atom within some molecule with other chemical elements, then we can evaluate the corresponding half-life 
of such an atom with respect to the electron capture. Currently, however, this problem can be solved only approximately, since there are quite a few difficulties in accurate computations of complex 
molecules as well as in actual experiments, since, e.g., the exact value of the constant $A$ in Eq.(\ref{eq1}) is not known. In other words, we cannot be sure that the experimental half-life mentioned 
above (53.60 days) corresponds to the electron capture from in ground $2^{1}S-$state of an isolated ${}^{7}$Be atom. In fact, it is not clear what chemical compounds were used (and at what conditions) 
to obtain this half-life. Very likely, we are dealing with some `averaged' value determined for a mixture of different molecules. It is clear that improving the overall experimental accuracy and purity of 
future experiments is critical. The accuracy of the future theoretical computations should be improved also. First of all, we need to focus on accurate expectation values of the electron-nucleus 
delta-function $\langle \delta({\bf r}_{eN}) \rangle$, rather than just accurate values of the total energy. Right now, we can only hope that in the future these problems can be solved and sorted out. Then 
the formula, Eq.(\ref{eq3}), can be used to determine the actual life-times of the ${}^{7}$Be atoms, which are included in different chemical compounds.       

\section{Conclusion}

We have considered the bound state properties of the negatively charged Li$^{-}$ ion in the ground $2^1S-$state. The same bound state properties are also determined for the ${}^{6}$Li$^{-}$ and 
${}^{7}$Li$^{-}$ ions with the finite nuclear masses and they are compared with the analogous properties of the neutral Li atom. Our analysis of the bound state properties of the Li$^{-}$ ion is of 
interest, since the formation of the negatively charged Li$^{-}$ ions plays an important role in the modern lithium and lithium-ion batteries. An extensive analysis of the bound state properties of the 
negatively charged Li$^{-}$ ion(s) has been performed. Expectation values of different properties determined in this study are sufficient for all current and future experimental needs. As follows 
from our calculations the Li$^{-}$ ion is a weakly-bound atomic system which has only one bound $2^{2}S-$state. The internal structure of this state is represented as a motion of one `almost free' 
electron in the field of a heavy atomic cluster which is the neutral Li atom in its ground $2^2S-$state. The computed expectation values of the bound state properties of the Li$^{-}$ ion in the 
$2^1S-$state and neutral Li atom in the $2^2S-$state do not contradict such a picture. Moreover, the whole internal structure of the Li$^{-}$ ion could be reconstructed to very good accuracy if we 
knew the model potential between an electron and neutral Li atom. This corresponds to the two-body approximation. An accurate reconstruction of such a model $e^{-}$-Li interaction potential should be a 
goal of future research. The same model potential can be used to obtain the cross-section of the elastic scattering (at relatively small energies) for the electron-lithium scattering.  

It should be mentioned that the negatively charged ${}^{6}$Li$^{-}$ ion is of interest for possible creation and observation of an unstable (three-electron) ${}^{4}$He$^{-}$ ion which is formed in one of 
the channels of the reaction of the ${}^{6}$Li$^{-}$ ion with slow neutrons, e.g., 
\begin{eqnarray}
 {}^6{\rm Li}^{-} + n = {}^4{\rm He}^{-} + {}^3{\rm H}^{+} + e^{-} + 4.785 \; MeV \; \; \; , \label{concl1}
\end{eqnarray}
Preliminary evaluations indicate that the probability of formation of the ${}^{4}$He$^{-}$ ion in this reaction is $\approx$ 0.02 \% - 0.04 \%. Nevertheless, this nuclear reaction of the 
${}^6$Li$^{-}$ ion with slow neutrons has a very large cross-section and it can be used to produce the negatively charged He$^{-}$ ion which is unstable and decays into the neutral He atom with the 
emission of one electron. Other approaches to create relatively large numbers of the negatively charged ${}^{4}$He$^{-}$ ions have failed. 

We also investigated the problem of experimental variations of the half-life of the beryllium-7 isotope placed in different chemical enviroment. Since the middle of 1930's this interesting problem has 
attracted a significant experimental and theoretical attention. It is shown here that the half-life of the beryllium-7 isotope in different chemical backgounds may vary by 3 \% - 6 \% (maximum). A central 
computational part of this problem is to determine to high accuracy the electron-nucleus delta-function of the Be-atom placed in different molecules, `quasi-metalic' alloys and other chemical compounds. 
The currently achieved accuracy is not sufficient to make accurate predictions of the half-life of the beryllium-7 atom in many molecules. Another part of the problem is to improve the overall purity 
and accuracy of current experiments performed with different molecules which include atoms of beryllium-7.  

\section{Acknowledgments}

We are grateful to Gerald F. Thomas and Prof. Will Williams (Smith Colledge, Northampton,  Massachusetts) for helpful discussions and useful references related to the beryllium-7 problem.

\begin{center}
  {\Large Appendix} \\
\end{center}

The expectation values $\langle {\bf p}_e \cdot {\bf p}_N \rangle$ and $\langle {\bf p}_e \cdot {\bf p}_e \rangle$ are not presented in Table I, since they are not truly independent from the $\langle 
\frac12 p^2_e \rangle$ and $\langle \frac12 p^2_N \rangle$ expectation values given in this Table. Indeed, for an arbitrary $K-$electron atom/ion the expectation values of the scalar products of the 
vectors of electron's momenta ${\bf p}_i$ ($i = 1, \ldots, K$) with each other and with the momentum of the nucleus ${\bf p}_N$ are simply related with the expectation values of the single-electron 
kinetic energy and kinetic energy of the nucleus:
\begin{eqnarray}
 & & \langle {\bf p}_i \cdot {\bf p}_j \rangle = \langle {\bf p}_1 \cdot {\bf p}_2 \rangle = \frac{2}{K (K - 1)} \Bigl[ \langle \frac12 p^2_N \rangle - 2 \langle \frac12 p^2_e \rangle \Bigr] 
 \; \; \; \label{App1} \\
 & & \langle {\bf p}_i \cdot {\bf p}_N \rangle = \langle {\bf p}_1 \cdot {\bf p}_N \rangle = - \frac{2}{K} \langle \frac12 p^2_N \rangle \; \; \; \; , \; \label{App2} 
\end{eqnarray}
where $K$ is total number of electrons, $\langle {\bf p}_i \cdot {\bf p}_j \rangle$ is the scalar product of the electron momenta of two atomic electrons (with indexes $i$ and $j$), while $\langle 
{\bf p}_i \cdot {\bf p}_N \rangle$ is the scalar product of the atomic nucleus and electron (with index $i$). Since the electron's indexes are arbitrary we can replace these scalar products by the 
$\langle {\bf p}_1 \cdot  {\bf p}_2 \rangle$ and $\langle {\bf p}_1 \cdot  {\bf p}_N \rangle$ values, respectively. These expectation values determine the kinematic correlations, or kinematic
interparticle correlations. In Eqs.(\ref{App1}) and (\ref{App2}) the notations $\langle \frac12 p^2_e \rangle$ and $\langle \frac12 p^2_N \rangle$ designate the single-electron kinetic energy and 
nucleus kinetic energy, respectively. It is interesting to note that the nuclear charge $Q$ is not included in these expressions. These equalities are obeyed for an arbitrary $K-$electron atom/ion, 
where $K \le Q$ (or, in general, $K \le Q + 1$). Therefore, there is no need to include the $\langle {\bf p}_1 \cdot {\bf p}_2 \rangle$ and $\langle {\bf p}_1 \cdot {\bf p}_N \rangle$ expectation 
values in Table I. For two-electron atomic systems $K = 2$ and equalities mentioned above take the well known form \cite{Fro2007}
\begin{equation}
 \langle {\bf p}_1 \cdot {\bf p}_2 \rangle = \langle \frac12 p^2_N \rangle - 2 \langle \frac12 p^2_e \rangle \; \; \; , \; \; \; \langle {\bf p}_e \cdot {\bf p}_N \rangle = \langle {\bf p}_1 
 \cdot {\bf p}_N \rangle = - \langle \frac12 p^2_N \rangle \; \; \; \; , \; \label{App3} 
\end{equation}
These identities are often used as a test in highly accurate, bound state computations of various two-electron atom/ions and arbitrary three-body systems.

\newpage


 \begin{table}[tbp]
   \caption{The expectation values of a nuber of electron-nuclear ($en$) and electron-electron ($ee$) properties in $a.u.$ of the ground $2S^{1}-$ and $2^2S-$states 
            of the of the Li$^{-}$ (${}^{\infty}$Li$^{-}$) ion and neutral Li (${}^{\infty}$Li) atom, respectively. The bound states of the negatively charged 
            ${}^{6}$Li$^{-}$ and ${}^{7}$Li$^{-}$ ions are also presented (in $a.u.)$}
     \begin{center}
     \begin{tabular}{| c | c | c | c | c | c | c | c |}
       \hline\hline          
 atom/ion  & state  & $\langle r^{-2}_{eN} \rangle$ & $\langle r^{-1}_{eN} \rangle$ & $\langle r_{eN} \rangle$ & $\langle r^2_{eN} \rangle$  & $\langle r^3_{eN} \rangle$ & $\langle r^4_{eN} \rangle$ \\
     \hline
 Li$^{-}$        & $2^1S$ & 7.568083    & 1.474722  & 2.902121    & 17.47139  & 139.161  & 1325.8 \\

 Li              & $2^2S$ & 10.08030323 & 1.90603751 & 1.66316156 & 6.1179832 & 30.86392 & 183.3014 \\
           \hline
 ${}^6$Li$^{-}$  & $2^1S$ & 7.566679    & 1.474584   & 2.902367   & 17.47412  & 139.191  & 1326.2 \\

 ${}^7$Li$^{-}$  & $2^1S$ & 7.566879    & 1.474604   & 2.902332   & 17.47373  & 139.187  & 1326.1 \\
       \hline\hline     
 atom/ion  & state & $\langle r^{-2}_{ee} \rangle$ & $\langle r^{-1}_{ee} \rangle$ & $\langle r_{ee} \rangle$ & $\langle r^2_{ee} \rangle$ &  $\langle r^3_{ee} \rangle$ &  $\langle r^4_{ee} \rangle$ \\
     \hline
 Li$^{-}$        & $2^1S$ & 0.747580  & 0.448973 & 5.116179  & 37.93169  & 349.203 & 3765.8 \\

 Li              & $2^2S$ & 1.460396634 & 0.73273820 & 2.88943970 & 12.2821625 & 64.021857 & 385.1388 \\
                 \hline
 ${}^6$Li$^{-}$  & $2^1S$ & 0.747456  & 0.448935 & 5.116595  & 37.93748  & 349.277 & 3766.9 \\

 ${}^7$Li$^{-}$  & $2^1S$ & 0.747473  & 0.448941 & 5.116536  & 37.93665  & 349.266 & 3766.7 \\
      \hline\hline             
 atom/ion  & state & $E$ & $\langle \frac12 p^2_{e} \rangle$ & $\langle \frac12 p^2_{N} \rangle$ & $\langle \delta_{eN} \rangle$ & $\langle \delta_{ee} \rangle$ &  $\langle \delta_{eee} \rangle$  \\ 
     \hline
 Li$^{-}$        & $2^1S$ & -7.50076754  & 1.8755627   & 7.8100020   & 3.42828   & 0.091265   & 0.0 \\

 Li              & $2^2S$ & -7.478060077 & 2.492687684 & 7.779902938 & 4.6074655 & 0.18165531 & 0.0 \\
                 \hline
 ${}^6$Li$^{-}$  & $2^1S$ & -7.50005515  & 1.8752077   & 7.8082534   & 3.42733   & 0.091241   & 0.0 \\

 ${}^7$Li$^{-}$  & $2^1S$ & -7.50015679  & 1.8752584   & 7.8085029   & 3.42747   & 0.091245   & 0.0 \\
     \hline\hline
  \end{tabular}
  \end{center}
   \end{table}
\end{document}